\documentclass[12pt]{article}

\newcommand{\sect}[1]{ \section{#1} }

\newcommand{\ve}{\left( \begin{array}{r}}
\newcommand{\ev}{\end{array} \right)}
\newcommand{\ar}{\left( \begin{array}{rr}}
\newcommand{\ra}{\end{array} \right)}
\newcommand{\arr}{\left( \begin{array}{rrrr}}
\newcommand{\arrr}{\left( \begin{array}{rrrrrr}}

\newcommand{\eqr}{\begin{eqnarray}}
\newcommand{\rqe}{\end{eqnarray}}
\newcommand{\eq}{\begin{equation}}
\newcommand{\qe}{\end{equation}}

\def\KK{{\rm I\kern -.2em  K}}
\def\NN{{\rm I\kern -.16em N}}
\def\RR{{\rm I\kern -.2em  R}}
\def\ZZ{\bf Z}
\def\ZZZ{{\small{\rm Z}\kern -.5em Z}}
\def\QQ{{\rm \kern .25em
             \vrule height1.4ex depth-.12ex width.06em\kern-.31em Q}}
\def\CC{{\rm \kern .25em
             \vrule height1.4ex depth-.12ex width.06em\kern-.31em C}}

\title{Localized intersections of M5-branes and four-dimensional
superconformal field theories}
\author{Ansar Fayyazuddin$^1$\footnote{email: fayyaz@binah.cc.brandeis.edu}
\quad and \quad
Douglas J. Smith$^2$\footnote{email: smith@physik.hu-berlin.de} }

\begin{document}

\maketitle

\begin{center}

{\em
$^1$Martin Fisher School of Physics\\
Brandeis University\\
Waltham, MA 02138, U.S.A. \\
\vspace{0.5cm}
$^2$Institut f\"ur Physik\\
Humboldt-Universit\"at\\
D-10115 Berlin\\
Germany
}

\end{center}

\vspace{1.4cm}

\begin{abstract}
We write supersymmetry preserving conditions for infinite
M5-branes intersecting on a (3+1)-dimensional space.  In contrast to
previously known solutions, these intersections are completely localized. 
We solve the equations
for a particular class of configurations which in
the near-horizon decoupling limit are dual to $N_f = 2N_c$
Seiberg-Witten superconformal field theories with gauge group SU($N$) and
generalisations to SU($N$)$^n$.  We also discuss the
relationship
to D3-branes in the presence of an $A_k$ singularity.

\end{abstract}

\vspace{-20cm}
\begin{flushright}
BRX TH-452\\
HUB-EP-99/11\\
hep-th/9902210\\
\end{flushright}

\thispagestyle{empty}

\newpage

\setcounter{page}{1}

\sect{Introduction}
A recent conjecture due to Maldacena \cite{Maldacena} asserts that gauge
theories realized as worldvolume theories of branes can be described
equivalently as gravity theories in the near-horizon
geometry due to these branes.  A precise correspondence between
problems in gauge theory and problems in gravity has been elucidated
in \cite{AdS_ref_GKP, AdS_ref_W}.  

One interesting field of application is four-dimensional
conformal field theories. Of particular importance is the ${\cal N}=4$
SU($N$) system realized by parallel D3-branes in type IIB string
theory.  Its near-horizon geometry is $AdS_5 \times S^5$. 
A number of other conformal field theories with less supersymmetry have been
constructed by employing various techniques
\cite{ks,lnv,bkv,fs,Kehagias:1998gn,afm,kw,gubser,Lopez:1998zf,kpw,klm,kt}.
The near-horizon geometry due to the branes
is of the form $AdS_5 \times M$ where $M$ is a compact manifold. 
The isometries of the near-horizon geometry are realized as global symmetries of
the world-volume gauge theory of the branes.
In particular, the $AdS$ factor is a signature of conformal invariance of the world-volume
gauge theory since its
isometry group is the conformal group of the field theory living at its boundary.

In this paper we study ${\cal N}=2$ superconformal field theories.
We first describe the conditions for
supersymmetry-preserving configurations of 5-branes in 11-dimensional
supergravity. This will allow us to write down the general form of the metric
describing 5-branes with worldvolume $\RR^4 \times \Sigma$ preserving at least
one quarter of the maximum supersymmetry. This corresponds to ${\cal N} = 2$
supersymmetry in 4 dimensions.
We will then show how a specific configuration
of this form reduces to a Hanany-Witten brane configuration \cite{HW} in
10-dimensional type IIA \cite{witn2},
realising a 4-dimensional conformal field theory. This configuration
is special in that it receives no non-perturbative corrections
and so one can safely describe the system in type IIA.  We are able to
solve the equations only when the NS5-branes are smeared in one
direction. This smearing is typical of known configurations of intersecting
branes, \cite{Gauntlett, Lu:1997hb, Stelle:1998xg} and references therein.
See also \cite{Itzhaki:1998uz, Hashimoto:1998ug, Yang:1999ze} for recent work
on localised intersections.
However, our method allows, in principle, the construction of fully
localised solutions.
We will take the
near-horizon limit of this supergravity solution which should be the
Maldacena dual of the conformal field theory. We show that the space is of the
form $AdS_5 \times M$ as expected.  The compact manifold $M$ has the isometry group
SO(3)$\times$U(1) matching the R-symmetry group of the field theory, there
is also an additional U(1) due to the smearing.  This last U(1) should
disappear in the fully localized solution.

In addition we write down explicitly the T-dual \cite{ovafa, Witten:1995zh}
(see also \cite{Klemm:1996bj, kls, smith}) geometry in Type IIB.
This geometry describes D3-branes in the presence of an $A_1$ ALE singularity.
This parallels the recent discussions of D3-branes at conifold singularities
\cite{kw, Lopez:1998zf} which is T-dual to NS5- and D4-brane configurations
\cite{conbrane_Uranga, conbrane_DM, conbrane_Unge} with $N=1$
supersymmetry in four dimensions.  

\section{Preservation of supersymmetry}
\label{SUSY_preservation}

In this section we will discuss the conditions on the metric and 4-form $F$
for a supersymmetry-preserving solution of 11-dimensional supergravity. It is
known that a configuration preserving some supersymmetry will always be a
solution of the classical equations of motion provided the Bianchi identity
and equations of motion for $F$ are satisfied. So we will start by analysing
the restrictions imposed by the requirement that we preserve (in our case
one quarter) supersymmetry. This is essentially a BPS condition and so
ensures that the equations of motion are satisfied. We will then supplement
the resulting conditions
with the additional constraints on $F$.

The SUSY variation of the gravitino Rarita-Schwinger field $\Psi_{M}$ is:

\begin{equation}
\delta_{\epsilon} \Psi_{M} = \left ( \partial_{M} +
        \frac{1}{4} \omega^a_{~bM}\hat{\Gamma}_a^{~b} +
        \frac{1}{144} \Gamma_M^{~~NPQR} F_{NPQR} -
        \frac{1}{18} \Gamma^{PQR} F_{MPQR} \right ) \epsilon
\label{SUSYvar}
\end{equation}
In order to find supersymmetric solutions we should choose the metric and the
4-form $F$ so that:
\begin{equation}
\delta_{\epsilon} \Psi_{M} = 0
\end{equation}
for some $\epsilon$ where in our conventions:
\begin{equation}
\hat{\Gamma}_{(10)}\epsilon \equiv \hat{\Gamma}_{0123456789}\epsilon
\end{equation}

We want a solution which corresponds to an M5-brane with worldvolume
$\RR^4 \times \Sigma$ with $\Sigma$ being a 2-cycle holomorphically embedded
in $\CC^2$ with complex coordinates $(v,s)$. I.e.\ $\Sigma$ is specified as
the zero of a holomorphic function $f(v,s)$. We can also consider several
5-branes by taking $\Sigma$ to be a sum of disconnected holomorphic 2-cycles.
The metric should be symmetric under 4d Lorentz transformations
and rotations in the three transverse directions.  If we pick a complex
structure in the $x_4,x_5,x_6,x_7$ directions and assume that the
metric is Hermitian in this subspace, the most general
metric consistent with the above symmetries is of the form:
\begin{equation}
ds^2 = H_1\eta_{\mu \nu}dx^{\mu}dx^{\nu} +
        2G_{m \overline{n}}dz^mdz^{\overline{n}} +
        H_2\delta_{\alpha \beta}dx^{\alpha}dx^{\beta}
\end{equation}
where (middle) Greek indices $\mu, \nu$ etc. correspond to $M = 0, 1, 2, 3$ 
and (beginning) Greek indices $\alpha, \beta$ etc. correspond to
$M = 8, 9, 10$. The two complex variables are defined by $z^v = v = x^4 + ix^5$
and $z^s = s = x^6 + ix^7$. $H_1, H_2$ and $G_{m \overline{n}}$ are arbitrary
functions of $v, \overline{v}, s, \overline{s}$ and
$r = \sqrt{\delta_{\alpha \beta}x^{\alpha}x^{\beta}}$.
For later convenience we will also define the rescaled complex metric:
\begin{equation}
g_{m \overline{n}} = H_1^{-1} G_{m \overline{n}}
\end{equation}

In order to recover Minkowski space far from any sources we
must have asymptotically:
\begin{eqnarray}
H_1 & \rightarrow & 1 \\
H_2 & \rightarrow & 1 \\
G_{v \overline{v}} & \rightarrow & \frac{1}{2} \label{bc_s} \\
G_{s \overline{s}} & \rightarrow & \frac{1}{2} \\
G_{v \overline{s}} & \rightarrow & 0 \\
G_{s \overline{v}} & \rightarrow & 0 \label{bc_e}
\end{eqnarray}

We now wish to analyse the condition for preserving supersymmetry. For a
5-brane with worldvolume $\RR^4 \times \Sigma$, the number of supersymmetries
preserved is given by the number of spinors $\epsilon$ which satisfy
\cite{Becker:1995kb, Fayyazuddin:1997by, Cvetic:1997yf, Fayyazuddin:1997ky}:
\begin{equation}
\sqrt{\det h} d\sigma^1d\sigma^2 \epsilon =
        \hat{\Gamma}_{0123}\Gamma_{m\overline{n}}
        (\partial_1Z^m\partial_2Z^{\overline{n}} - \partial_1Z^m\partial_2Z^{\overline{n}}) d\sigma^1d\sigma^2 \epsilon
\end{equation}
where $h_{ij}$ is the pullback of the spacetime metric $G_{m\overline{n}}$
onto $\Sigma$, $\sigma^i$ are the two (real) worldvolume coordinates on
$\Sigma$ and $Z^m(\sigma^1, \sigma^2)$ are the target space coordinates of
$\Sigma$. Now on $\Sigma$ we have $f(Z^v, Z^s) = 0$ so:
\begin{equation}
\partial_iZ^s = -\partial_vf(\partial_sf)^{-1}\partial_iZ^v
\end{equation}
We can now evaluate the determinant of
$h_{ij} = \partial_{(i}Z^m\partial_{j)}Z^{\overline{n}} G_{m\overline{n}}$ as:
\begin{equation}
\sqrt{\det h} = \pm i(\partial_{[1}Z^v\partial_{2]}Z^{\overline{v}})
        \left ( G_{v\overline{v}} -
        \overline{ \left ( \frac{\partial_vf}{\partial_sf} \right ) }G_{v\overline{s}} -
        \frac{\partial_vf}{\partial_sf}G_{s\overline{v}} +
        \left | \frac{\partial_vf}{\partial_sf} \right |^2 G_{s\overline{s}}
        \right )
\end{equation}
so the equation for preservation of supersymmetry becomes:
\begin{eqnarray}
\hat{\Gamma}_{0123} \left (\Gamma_{v\overline{v}} -
        \overline{\left ( \frac{\partial_vf}{\partial_sf} \right ) }\Gamma_{v\overline{s}} -
        \frac{\partial_vf}{\partial_sf}\Gamma_{s\overline{v}} +
        \left | \frac{\partial_vf}{\partial_sf} \right |^2 \Gamma_{s\overline{s}}
        \right ) \epsilon & & \nonumber \\
= ic \left ( G_{v\overline{v}} -
        \overline{\left ( \frac{\partial_vf}{\partial_sf} \right ) }G_{v\overline{s}} -
        \frac{\partial_vf}{\partial_sf}G_{s\overline{v}} +
        \left | \frac{\partial_vf}{\partial_sf} \right |^2 G_{s\overline{s}}
        \right ) \epsilon & &
\end{eqnarray}
where $c = \pm 1$ corresponding to a brane or anti-brane.
So we end up with the conditions that the spinor $\epsilon$ must satisfy:
\begin{equation}
\hat{\Gamma}_{0123}\Gamma_{m\overline{n}}\epsilon = icG_{m\overline{n}}\epsilon
\end{equation}
or equivalently
\begin{equation}
\hat{\Gamma}_{0123a\overline{b}} \epsilon = ic\delta_{a\overline{b}} \epsilon
\end{equation}
where
$\delta_{a\overline{b}}$ is the flat space metric with
$\delta_{1\overline{1}} = \frac{1}{2}$ in our conventions
\footnote{We use $a, b$ etc. with values 1 and 2 for the flat space complex
coordinates while $m, n$ etc. with values $v$ and $s$ are the curved space
complex coordinate indices.}.
This restriction on $\epsilon$ means that the solution will preserve
$\frac{1}{4}$ of the supersymmetry which corresponds to ${\cal N} = 2$ in
four dimensions.

Using these equations we can express the terms on the right hand side of
equation~(\ref{SUSYvar}) in terms of a sum of linearly-independent products
of gamma matrices
acting on $\epsilon$. The coefficients of these terms must then vanish and
this will give us a set of equations which must be satisfied for a solution
preserving $\frac{1}{4}$ of the supersymmetry.



It is straightforward to express equation~(\ref{SUSYvar}) as a sum of
independent products of $\Gamma$-matrices and we will only present the
final results here. Requiring the coefficient of each
term to vanish we get a large number of
equations which can be reduced to the following set of independent equations:
\begin{eqnarray}
H_1^2 & = & H_2^{-1} \\
H_2 & = & (4g)^{\frac{2}{3}} \equiv H^{\frac{2}{3}} \\
F_{m \overline{n} \alpha \beta} & = &
        \frac{ic}{2} \epsilon_{\alpha \beta \gamma} \partial_{\gamma} g_{m \overline{n}} \label{F_g_start} \\
F_{m89(10)} & = & -\frac{ic}{2} \partial_m H \\
F_{\overline{m}89(10)} & = & \frac{ic}{2} \partial_{\overline{m}} H \label{F_g_end} \\
\partial_v g_{s \overline{n}} & = & \partial_s g_{v \overline{n}} \\
\partial_{\overline{v}} g_{m \overline{s}} & = &
        \partial_{\overline{s}} g_{m \overline{v}}
\end{eqnarray}
with all other types of components of $F$ vanishing.
The last two equations are simply the statement that $g_{m\overline{n}}$ is
a K\"ahler metric. Clearly the full metric and 4-form $F$ are determined
explicitly by the K\"ahler metric.

Now to ensure that the equations of motion of 11-dimensional supergravity are
satisfied we only need to satisfy the Bianchi identity and equations of motion
for $F$. This will give us the equations to determine $g_{m \overline{n}}$.
We note that $F \wedge F = 0$ for above $F$.
Using the above relations it is easy to verify that the Bianchi identity for
the dual 7-form $*F$:
\begin{equation}
d*F = 0
\end{equation}
is automatically satisfied and we have the equations of
motion with source term:
\begin{equation}
dF = J_{m \overline{n}} dz^m \wedge dz^{\overline{n}} \wedge
        dx^8 \wedge dx^9 \wedge dx^{10},
\end{equation}
where $J$ is the source for $*F$, leading to the following equations:
\begin{equation}
J_{m \overline{n}} = \frac{ic}{2} \left ( 4 \partial_m\partial_{\overline{n}} (2g) + \partial_{\gamma}\partial_{\gamma} g_{m\overline{n}} \right )
\label{source}
\end{equation}

\subsection{Simple example}

We can now check that we reproduce the solution for a single M5-brane with
worldvolume 012345. Let
\begin{eqnarray}
g_{v \overline{v}} & = & \frac{1}{2} \\
g_{s \overline{s}} & = & \frac{1}{2}H \\
g_{v \overline{s}} & = & 0 \\
g_{s \overline{v}} & = & 0
\end{eqnarray}
Then equations~(\ref{source}) reduce to:
\begin{equation}
J_{s \overline{s}} = \frac{ic}{4} (4\partial_s\partial_{\overline{s}} +
        \partial_{\gamma}\partial_{\gamma} ) H
\end{equation}
where the source term is just a delta function defining the position of
the M5-brane, normalised as:
\begin{equation}
J_{s \overline{s}} = -ic2\pi^3l_{p}^{3}\delta^{(3)}(r)\delta^{(2)}(s)
\end{equation}
So H is a solution of the 5-dimensional Laplace equation with a source at the
position of the 5-brane. So we find:
\begin{equation}
H = 1 + \frac{\pi l_p^3}{r_{(5)}^3}
\end{equation}
where $r_{(5)}$ is the radial coordinate in the 5 dimensions
transverse to the 5-brane. It is easy to see this reproduces the solution
given in \cite{Gauntlett, 5brane_solution}. Of course this solution will
preserve $\frac{1}{2}$ of the full supersymmetry since it describes only
one flat M5-brane.

\subsection{Summary of general solution}

What we have shown is that given a K\"ahler metric $g_{m \overline{n}}$ on a
complex 2-dimensional space which satisfies the source equations:
\begin{equation}
J_{m \overline{n}} = \frac{ic}{2} \left ( 4 \partial_m\partial_{\overline{n}}
 (2g) + \partial_{\gamma}\partial_{\gamma} g_{m\overline{n}} \right )
\end{equation}
there exists a solution to the equations of motion of 11-dimensional
supergravity preserving at least one quarter of the full supersymmetry. The
4-form $F$ is defined explicitly in terms of $g_{m \overline{n}}$ by
equations~(\ref{F_g_start})--(\ref{F_g_end}). The metric is given by:
\begin{equation}
ds^2 = H^{-\frac{1}{3}}\eta_{\mu \nu}dx^{\mu}dx^{\nu} +
        2H^{-\frac{1}{3}}g_{m\overline{n}}dz^mdz^{\overline{n}} +
        H^{\frac{2}{3}}\delta_{\alpha \beta}dx^{\alpha}dx^{\beta}
\label{soln_11}
\end{equation}
where:
\begin{equation}
H = 4g =
   4(g_{v\overline{v}}g_{s\overline{s}} - g_{v\overline{s}}g_{s\overline{v}})
\end{equation}
The boundary conditions, equations~(\ref{bc_s})--(\ref{bc_e}), for Minkowski
space at infinity become simply:
\begin{eqnarray}
g_{v \overline{v}} & \rightarrow & \frac{1}{2} \label{bc_start} \\
g_{s \overline{s}} & \rightarrow & \frac{1}{2} \\
g_{v \overline{s}} & \rightarrow & 0 \\
g_{s \overline{v}} & \rightarrow & 0 \label{bc_end}
\end{eqnarray}

We can combine the K\"ahler conditions and the source equations into
equations determining the K\"ahler potential $K$. The metric is
defined as:
\begin{equation}
g_{m \overline{n}} = \partial_m \partial_{\overline{n}} K
\end{equation}
and the K\"ahler potential is determined by:
\begin{equation}
\frac{ic}{2} \partial_{m}\partial_{\overline n}(8g(K) + \partial_{\gamma} \partial_{\gamma} K) = J_{m \overline{n}}.
\label{K_pot}
\end{equation}
These equations are similar
to the Monge-Ampere equation and the general solution is not known.
It would be possible to find solutions numerically but we will not
pursue that approach here. It is also possible to solve asymptotically
around flat Minkowski spacetime which would be valid far from the
branes. However, we are interested in the near-horizon limit and in the next
section we will show the solution in this limit for a special configuration
of branes corresponding to the description of conformal QCD.

We note that even without solving equations~(\ref{K_pot}), the general form
of the solution exhibits the R-symmetry group appropriate for ${\cal N} = 2$
Yang-Mills in four dimensions. The overall transverse 3-space clearly has an
SO(3) isometry group and there is also a U(1) isometry for the
superconformal case, since $K$ depends on $v$ only through its absolute value
 in this case, as we see below. 
So as expected the global R-symmetry of
the gauge theory is manifested as the group of isometries of the metric.

It can be checked that our solution generalises previous solutions for
M5-branes intersecting orthogonally with a common (3+1)-dimensional
worldvolume (and related configurations in lower dimensions.)
Indeed if we choose $g_{v \overline{s}} = g_{s \overline{v}} = 0$
then the K\"ahler conditions tell us that $g_{v \overline{v}}$ is independent
of $s$ and $\overline{s}$ while $g_{s \overline{s}}$ is independent of $v$
and $\overline{v}$. It is then easy to show that equations~(\ref{source})
are equivalent to the curved space Laplace equations in \cite{Tseytlin:1996bh}.
It is important to also keep in mind the allowed coordinate dependence of the
functions, imposed in our case by the K\"ahler condition. This is equivalent
to the additional constraints on the `harmonic' functions, derived in
\cite{Lu:1998mi}. It is easy to see that this ansatz is too restictive to allow
solutions for fully localised intersecting branes -- at least one of the branes
must be smeared in some directions. Our general solution for this type of
brane configuration does allow fully localised branes to intersect. The
important generalisation seems to be the inclusion of non-diagonal terms in
the metric, i.e. we must have non-zero $g_{v \overline{s}}$ and
$g_{s \overline{v}}$. The explicit construction of such a solution requires
the solution of equations~(\ref{K_pot}) with appropriate fully localised
sources. Unfortunately we have so far been unable to find such a solution
and this matter is currently under investigation \cite{danda2}.

\section{Near-horizon limit of solution in 10 dimensions}
\label{D4NS5solution}

We want to consider the case of 
2 parallel NS5-branes with worldvolume 012345 and $N$ infinite D4-branes with
worldvolume 01236 in type IIA. This is the Hanany-Witten construction of
four-dimensional ${\cal N} =2$ SU($N$) gauge theory with $2N$ hypermultiplets
in the fundamental representation.  This is a conformal field theory.
Since the D4-branes are infinite they
can be thought of as aligned D4-branes ending on either side of the NS5-branes.  Such
a set-up does not result in bending of the branes since the force exerted by a D4-brane
on one side of a NS5-brane is exactly compensated by the aligned D4-brane on the other
side of the NS5-brane.  This fact has the important consequence that when we lift 
this configuration to M-theory it receives no non-perturbative corrections
\footnote{To be more precise we mean that there are no corrections to the
brane configuration. In field theory one observes \cite{Dorey} that the
coupling constant and moduli are in fact redefined compared to their
classical values. These redefinitions can be understood as loop and instanton
effects. In the brane picture we use the Seiberg-Witten
curve to determine the geometry. However, as observed in \cite{Dorey}
the curve itself does not change; only the precise relationship
between microscopic and low-energy quantities is modified. So in this
sense there are no non-perturbative effects.}.

Clearly this 10-dimensional background can be obtained from an 11-dimensional
background of M5-branes intersecting so as to preserve one quarter
supersymmetry,
as discussed in section~\ref{SUSY_preservation}.  We can lift the Hanany-Witten
configuration to two M5-branes with world-volume $012345$ localized at $t=t_{1}, r=0$
and $t=t_{2}, r=0$ respectively; and $N$ M5-branes
with world-volume $012367$ localized at $v=0, r=0$.
The Riemann surface $\Sigma$ mentioned in the previous section 
is easy to construct using the methods of \cite{witn2}.  
According to \cite{witn2} the Riemann surface can be represented as :
\begin{equation}
A(v)t^2 + B(v)t + C(v)=0 \hspace{1cm} (\hbox{with $t=\exp -s/R_7$}).
\end{equation}
Since we are compactifying M-theory on a circle of radius $R_7$ to give
type IIA, we use the single-valued coordinate $t$ instead of $s$.

Now since for arbitrary $t$ we are on the world-volume of the M5-brane only if $v=v_i$,
i.e. at the locations of the D4-branes,
we conclude that $A=\alpha B= \beta C=\prod_{i=1}^{N}(v-v_i)$.
The constants $\alpha, \beta$ are fixed by requiring that $t^2 + t/\alpha +1/\beta =0$
has solutions at $t=t_{1,2}$, the positions of the two NS5-branes.  We will be interested in the
case of maximal gauge symmetry enhancement when the D4-branes coincide, i.e. $v_i = 0$.
The Riemann surface $\Sigma$ then splits up into three holomorphic cycles:
i) $v=0$, $t$ arbitrary, ii) $t=t_{1}$, $v$ arbitrary, iii) $t=t_2$ with $v$ arbitrary.
Thus the lift of the type IIA Hanany-Witten configuration is 2 M5-branes localized at
$t=t_{1,2}$ and $N$ M5-branes localized at $v=0$ with the overall transverse coordinates
$x_\alpha =0$.  
In the following section we will show in detail
how to use these results to construct the 10-dimensional metric.

Since we wish to make contact with Maldacena's conjecture, we will really
be interested in the near-horizon limit of this solution.  Here we use a 
naive interpretation of near-horizon
as simply meaning that the harmonic functions approach zero at infinity instead
of 1.
This limit will in fact
allow us to find a solution where the $N$ D4-branes are fully localised. We
will then show that in the appropriate limit the metric contains an $AdS_5$
factor as is required for the equivalence via Maldacena's duality to a
4-dimensional conformal field theory.

\subsection{11 to 10 dimensions}

We have 2 M5-branes with worldvolume 012345 and $N$ with worldvolume 012367.
We will compactify the 7 direction on a circle to produce the required
Hanany-Witten brane construction in type IIA. For simplicity we will assume
translation invariance in the 6 direction as well as the 7 direction, i.e. we
will have the $N$ M5-branes fully localised but the 2 M5-branes will be
`smeared' in the 6 and 7 directions (or $s$ and $\overline{s}$ directions.)
The standard relation between 11-dimensional supergravity and 10-dimensional
type IIA `string-frame' metrics is \cite{Witten:1995ex, Bergshoeff:1995as}:
\begin{equation}
ds_{11}^2 = e^{-\frac{2\phi}{3}}ds_{10}^2 +
                e^{\frac{4\phi}{3}}(dx_7 + A_idx^i)^2
\end{equation}
Using the form of the 11-dimensional solution, equation~(\ref{soln_11}),
we find:
\begin{eqnarray}
e^{\frac{4\phi}{3}} & = & 2H^{-\frac{1}{3}}g_{s\overline{s}} \\
A_4 & = & \frac{-i}{2g_{s\overline{s}}}(g_{v\overline{s}}-g_{s\overline{v}}) \\
A_5 & = & \frac{1}{2g_{s\overline{s}}}(g_{v\overline{s}}+g_{s\overline{v}})
\end{eqnarray}
with all other components of $A_i$ vanishing.
So the general form of the 10-dimensional metric is:
\begin{eqnarray}
ds_{10}^2 & = &
 \frac{1}{\sqrt{2}}g^{-\frac{1}{2}}g_{s\overline{s}}^{\frac{1}{2}}\eta_{\mu\nu}dx^{\mu}dx^{\nu}
 + \sqrt{2}g^{\frac{1}{2}}g_{s\overline{s}}^{-\frac{1}{2}}dvd\overline{v} +
 2\sqrt{2}g^{\frac{1}{2}}g_{s\overline{s}}^{\frac{1}{2}}\delta_{\alpha\beta}dx^{\alpha}dx^{\beta} +
\nonumber \\
 & & \sqrt{2}g^{-\frac{1}{2}}g_{s\overline{s}}^{\frac{3}{2}} \left ( dx_6 +
  \frac{1}{2g_{s\overline{s}}}(g_{s\overline{v}}+g_{v\overline{s}})dx_5 +
  \frac{i}{2g_{s\overline{s}}}(g_{s\overline{v}}-g_{v\overline{s}})dx_4
  \right )^2
\end{eqnarray}
The 10-dimensional 3- and 4-form field strengths are given in terms of the
11-dimensional 4-form by \cite{Bergshoeff:1995as}:
\begin{eqnarray}
F^{10-d}_{MNPQ} & = & F^{11-d}_{MNPQ} \\
H_{MNP} & = & 2F^{11-d}_{MNP7}
\end{eqnarray}

\subsection{Ansatz for solution}

We will now assume that the metric $g_{m\overline{n}}$ is of the form:
\begin{eqnarray}
g_{s\overline{s}} & = & \frac{q_1}{r} \\
g_{v\overline{v}} & = & g_{v\overline{v}}(|v|,r) \\
g_{s\overline{v}} & = & g_{v\overline{s}} = 0
\end{eqnarray}
where $r = \sqrt{\delta_{\alpha\beta}x^{\alpha}x^{\beta}}$ is the radial
coordinate in the overall transverse 3-dimensional
space and $q_1$ is a constant. Note that this ansatz guarantees that
$g_{m\overline{n}}$ is K\"ahler.

So now we must find a solution to the source equations~(\ref{source}). The
sources will be a delta function in the 3 overall transverse dimensions
corresponding to the 2 smeared 5-branes and a delta function in the
5-dimensional space transverse to the $N$ 5-branes. So we see that
equations~(\ref{source}) are satisfied, up to normalisation of $q_1$, by our
ansatz except for the case of
the $v\overline{v}$ components.
For the $v\overline{v}$ components
we have (using a coordinate transformation used in \cite{Itzhaki:1998uz}):
\begin{eqnarray}
8 \partial_v\partial_{\overline{v}} g +
\partial_{\gamma}\partial_{\gamma} g_{v\overline{v}} & = &
\frac{2q_1}{r}\frac{1}{|v|}\partial_{|v|}(|v|\partial_{|v|}g_{v\overline{v}}) +
\frac{1}{r^2}\partial_r(r^2\partial_rg_{v\overline{v}}) \nonumber \\
 & = & \frac{16q_1^2}{\rho^2}\frac{1}{|v|}\partial_{|v|}(|v|\partial_{|v|}g_{v\overline{v}}) +
 \frac{16q_1^2}{\rho^5}\partial_{\rho}(\rho^3\partial_{\rho}g_{v\overline{v}})
 \nonumber \\
 & = & \frac{16q_1^2}{\rho^2}\frac{1}{R^5}\partial_R(R^5\partial_Rg_{v\overline{v}})
\label{delta_6}
 \end{eqnarray}
where $\rho = \sqrt{8q_1r}$ and $R^2 = \rho^2 + |v|^2$ and in the last step we
have assumed that $g_{v\overline{v}}$ is only a function of $R$.

So we can now see that we have a solution for $g_{v\overline{v}}$:
\begin{equation}
g_{v\overline{v}} = \frac{q_2}{R^4} = \frac{q_2}{(8q_1r + |v|^2)^2}
\end{equation}
where $q_2$ is a constant. We will now fix the constants $q_1$ and $q_2$.

In M-theory the normalization of the source equations is as
follows:
\begin{eqnarray}
\nabla^2 g_{s\overline{s}} + 8\partial_{s}\partial_{\overline{s}}g
& = & -4 \pi^3l_{p}^{3}\delta^{(3)}(r) \left ( \delta^{(2)}(s-L/2)
+\delta^{(2)}(s+L/2) \right ) \nonumber \\
\nabla^2 g_{v\overline{v}} + 8\partial_{v}\partial_{\overline{v}}g
& = & -N4\pi^3l_{p}^{3}\delta^{(3)}(r)\delta^{(2)}(v) \nonumber \\
\nabla^2 g_{v\overline{s}} + 8\partial_{v}\partial_{\overline{s}}g
& = & 0
\end{eqnarray}
We can reduce the solution down to 10-dimensions by compactifying
$x_7$ with radius $R_{7}$.  When $R_7$ is small enough we can
smear branes in the $x_7$ direction and replace the delta
function by $1/2\pi R_{7}$ \footnote{This procedure gives the
same result as the Poisson summation of the harmonic function
given in \cite{IMSY}.}.  Using the relation
$l_{p}^{3} = R_{7}\alpha '= g_{s}(\alpha ')^{3/2}$ we get:
\begin{eqnarray}
\nabla^2 g_{s\overline{s}} + 2\partial_{6}\partial_{6}g
& = & -2\pi^2\alpha ' \delta^{(3)}(r) \Big ( \delta(x_6-L/2)
+\delta(x_6 +L/2) \Big ) \nonumber \\
\nabla^2 g_{v\overline{v}} + 8\partial_{v}\partial_{\overline{v}}g
& = & -N4\pi^3g_{s}(\alpha ')^{3/2} \delta^{(3)}(r)\delta^{(2)}(v) \nonumber \\
\nabla^2 g_{v\overline{s}} + 4\partial_{v}\partial_{6}g
& = & 0.
\end{eqnarray}

Finally, we smear the 2 NS5-branes in the $x_{6}$ direction.
We compactify the $x_6$ direction with radius $\sqrt{\alpha '}/c_1$.
Smearing in the $6$ direction,  we can replace the delta functions by 
$c_1 / 2\pi \sqrt{\alpha '}$. More generally, if we have $n$ NS5-branes the
source equation becomes:
\begin{equation}
\nabla^2 g_{s\overline{s}} + 2\partial_{6}\partial_{6}g
 =  -\pi nc_1 (\alpha ')^{1/2}\delta^{(3)}(r).
\end{equation}

We can now fix the normalization
\begin{equation}
q_{1}= n(\alpha ')^{1/2} c_1 / 4
\end{equation}
and after inserting appropriate volume factors to convert the 6-dimensional
delta function at $R = 0$ in equation~(\ref{delta_6}) to delta functions in
$r$ and $v$:
\begin{eqnarray}
q_{2} & = & 8\pi g_{s}(\alpha ')^{3/2}Nq_1 \\
      & = & 2\pi g_s (\alpha ')^2 nNc_1
\end{eqnarray} 

\subsection{Decoupling limit}

The near-horizon limit is one in which field theory quantities are held
fixed while decoupling the bulk degrees of freedom.  It is natural
to express the metric in terms of variables appropriate to this limit.

In this system of D4-branes and NS5-branes we would like to preserve
the masses of field theory excitations and the field theory coupling constant.
The moduli space of the field theory is parameterized by a complex adjoint Higgs 
field.  On the supergravity side the Higgs expectation values are given
by the positions of the D4-branes in $v$.  The mass of a string stretched
between two D4-branes is $|w|= |v| /\alpha '$ where $|v|$ is the coordinate
distance between the branes.  These states are ``electrically'' charged.
There are also magnetically charged states which are described by
D2-branes stretched between D4-branes.  They have mass: 
\begin{equation}
m = \frac{|v| L}{g_s(\alpha ')^{3/2}}
= \frac{|v|}{g_{YM}^{2}\alpha '} = \frac{|w|}{g_{YM}^{2}},
\end{equation}
where $L$ is the coordinate distance between the two NS5-branes.
Thus in the $\alpha '\rightarrow 0$ limit if we keep $w=v/\alpha '$ and the Yang-Mills
coupling constant $g_{YM}$ fixed, the field theory states have finite mass. 
  
The field theory living on the D4-branes has only $v$ as a modulus and the field
theory excitations only determine the scaling of $v$.  The
remaining coordinates $r,x_{6}$ in the decoupling limit have their scalings
determined by keeping excitations on the NS5-brane and the Yang-Mills coupling
constant fixed.  The NS5-branes have D2-branes ending on them.  The ends
of the D2-brane appear as strings in the NS5-brane's world volume
theory with tension $r/g_s{\alpha '}^{3/2}$.  In the decoupling limit we
take $\alpha '\rightarrow 0$ while keeping $r/g_s{\alpha '}^{3/2}$ fixed.

Lastly the scaling of $x_6$ can be determined as follows in the decoupling
limit.  The Yang-Mills coupling constant is given by 
$g_{YM}^{-2}= L/g_{s}(\alpha ')^{1/2}$, where $L$ is the coordinate
distance between the two NS5-branes.  Thus to keep the coupling
constant fixed in the field theory we have to take the limit $\alpha '\rightarrow 0$
while keeping $x_{6}/g_s{\alpha '}^{1/2}$ fixed. 

With $n$ NS5-branes the gauge group is SU($N$)$^n$.
The gauge coupling of each SU($N$) group is given by the separation, $L$, of
two consecutive NS5-branes in the $x_6$ direction.
Since we have smeared the NS5-branes around the circle, they should be
considered to be equally spaced. This circle has radius $\sqrt{\alpha '} / c_1$
so $L = 2 \pi \sqrt{\alpha '} / (n c_1)$. So we have:
\begin{equation}
\frac{1}{g_{YM}^2} = \frac{L}{g_s \sqrt{\alpha '}} = \frac{2 \pi}{n g_s c_1}
\end{equation}
So we should keep $n g_s c_1$ fixed or if we take the large $N$ limit with
fixed 't Hooft coupling, we should keep $n N g_s c_1$ fixed.

So we will define variables $u, w$ and $t$ with dimensions of mass and a
dimensionless angular variable $\Phi$:
\begin{eqnarray}
u & = & \frac{R}{\alpha'} \\
w & = & \frac{v}{\alpha'} \\
t^2 & = & \frac{r}{g_s\alpha'^{\frac{3}{2}}} \\
\Phi & = & \frac{c_1 x_6}{\sqrt{\alpha'}}
\end{eqnarray}
In terms of these variables the non-zero components of the complex metric are:
\begin{eqnarray}
g_{v \overline{v}} & = & \frac{2\pi g_s nNc_1}{(\alpha ')^2 u^4} \\
g_{s \overline{s}} & = & \frac{nc_1}{4g_s \alpha ' t^2}
\end{eqnarray}
Then we can express the 10-dimensional metric in terms of these variables as:
\begin{eqnarray}
\frac{1}{\alpha'}ds_{10}^2 & = &
        u^2Q^{-\frac{1}{2}}\eta_{\mu\nu}dx^{\mu}dx^{\nu} +
        u^{-2}Q^{\frac{1}{2}}dwd\overline{w} +
\nonumber \\
 & & u^2Q^{-\frac{1}{2}}\frac{nc_1}{2\alpha ' c_1t^2}d\Phi^2 +
    u^{-2}Q^{\frac{1}{2}}\frac{ng_sc_1}{2}(4dt^2 + t^2d\Omega^2_2)
\end{eqnarray}
where we have defined:
\begin{equation}
Q = 4 \pi g_s nNc_1
\end{equation}
We now define the dimensionless variables $\eta$ and $\theta$ with the ranges
$0 \le \eta \le \frac{\pi}{2}$ and $0 \le \theta < 2\pi$:
\begin{eqnarray}
w & = & |w|e^{i\theta} \\
|w| & = & u \sin\eta.
\end{eqnarray}
Then we have:
\begin{equation}
u^2 \cos^2 \eta = 2nc_1g_s t^2.
\end{equation}
We also write $d\Omega_2^2$ in terms of angular variables
$0 \le \omega \le \pi$ and $0 \le \sigma < 2 \pi$ as:
\begin{equation}
d\Omega_2^2 = d\omega^2 + \sin^2 \omega d\sigma^2
\end{equation}
We can now rewrite the metric in a form where the $AdS_5$ factor is explicit:
\begin{eqnarray}
\frac{1}{\alpha'}ds_{10}^2 & = &
        \left ( \frac{u^2}{\sqrt{Q}}\eta_{\mu\nu}dx^{\mu}dx^{\nu} +
        \frac{\sqrt{Q}}{u^2}du^2 \right ) +
        \left ( \sqrt{Q}d\eta^2 + \sqrt{Q} \sin^2\eta d\theta^2 + \right.
\nonumber \\
 & & \left. \frac{n^2}{\sqrt{Q}\cos^2\eta}d\Phi^2 +
     \frac{\sqrt{Q}}{4}\cos^2\eta (d\omega^2 + \sin^2 \omega d\sigma^2) \right)
\label{metric}
\end{eqnarray}
It is straightforward to find the non-zero 3- and 4-form field strengths. They
are given by:
\begin{eqnarray}
H_{\Phi \omega \sigma} & = & -\frac{c}{2} n \alpha ' \sin \omega \\
F_{\eta \theta \omega \sigma} & = & -2 \pi c g_s N (\alpha ')^{3/2} \sin \eta
			\cos^3 \eta \sin \omega
\end{eqnarray}
Since $H = dB$ we can calculate the non-trivial part of the B-field. We find:
\begin{equation}
B_{\Phi \sigma} = -\frac{c}{2} n \alpha ' (\cos \omega + c_0)
\end{equation}
where we have chosen to keep the constant of integration $c_0$.

\section{T-Duality}

Using the T-duality rules \cite{Buscher1, Buscher2, Bergshoeff:1995as}
we can T-dualise the metric
(\ref{metric}) to a type IIB metric for D3-branes at an $A_{n-1}$ singularity.
The near horizon limit is described by type IIB supergravity on the space
$AdS_5 \times (S^5 / \ZZ_n)$. The AdS/CFT correspondence in this case has been
discussed in \cite{ks, lnv}.
We perform the T-duality in the 6 direction which we have already considered
to be compact, parameterised by the angular coordinate $\Phi$.
As described in \cite{ovafa}, T-duality transverse to $n$
NS5-branes transforms them to $n$ Kaluza-Klein 5-branes which are equivalent
to an $n$-centred Taub-NUT space in the transverse directions. When the centres
coincide as in this case there is an $A_{n-1}$ singularity and at low
energies we can simply describe the space as a $\CC^2 / \ZZ_n$ orbifold. In
our case where the D4-branes fill the 6 direction we will end up
with D3-branes at the orbifold fixed point. However this T-duality can still
be performed for general Hanany-Witten setups with D4-branes ending on
NS5-branes. In this case the D4-branes will T-dualise to fractional D3-branes
at the orbifold fixed point \cite{kls, smith}. So the problem of finding the
supergravity solution for a general (non-conformal) Hanany-Witten setup is
equivalent to finding the metric for fractional D3-branes at an orbifold
fixed point.

We now briefly describe how to write down the metric for D3-branes
in the presence of an $A_{k-1}$ ALE singularity. An ALE space
with an $A_{k-1}$ singularity can be defined as a $\CC^{2} / \ZZ_{k}$
orbifold, where the action of the $\ZZ_{k}$ is given as follows:
\begin{eqnarray}
z_{1} & \rightarrow & \exp{(i\frac{2\pi}{k})}z_{1} \nonumber \\
z_{2} & \rightarrow & \exp{(-i\frac{2\pi}{k})}z_{2} \label{ale}.
\end{eqnarray}
The coordinates $z_1, z_2$ are covering space coordinates of the
ALE space, and in these coordinates the metric of the ALE space
is flat.  Thus in the covering space coordinates one can easily
write down the solution for D3-branes \cite{fs,afm}.  In fact, 
it is identical to the usual metric \cite{horstro} except now 
one has to introduce images of the D3-branes due to the identification 
(\ref{ale}).  To have $N$ D3-branes in the presence of the ALE singularity
one has to introduce $Nk$ D3-branes in the covering space, where $N(k-1)$
of them are images.  The metric for all $N$ D3-branes localized at
the origin is given by:  
\begin{equation}
ds^2 = f^{-1/2}dx_{\parallel}^{2} + f^{1/2} \left ( |dv|^2 +
|dz_1|^2 + |dz_2|^2 \right)
\label{covmet}
\end{equation}
where
\begin{equation}
f = 1 +\frac{4\pi g_skN{\alpha '}^2}{r^4}.
\end{equation}
We can now change variables from the covering space to single-valued
variables as in \cite{Itzhaki:1998uz}:
\begin{eqnarray}
\rho^2 & = & |z_1|^2 + |z_2|^2 \nonumber \\
z_1  & = &\rho\cos{\frac{\theta}{2}}e^{i\frac{\delta}{k}} \\
z_2  & = & \rho\sin{\frac{\theta}{2}}e^{i\eta - i\frac{\delta}{k}}
\end{eqnarray}
with $0\le\theta\le\pi$, $0\le\delta <2\pi$ and $0\le\eta <2\pi$.
We also introduce
\begin{eqnarray}
\rho &=& r\sin{\gamma} \nonumber\\
v &=& r\cos{\gamma}e^{i\phi}
\end{eqnarray}
with $0\le\gamma\le\pi /2$, $0\le\phi <2\pi$. We can now take the
near-horizon limit $\alpha '\rightarrow 0$  with 
 $u = r/\alpha '$ held fixed where we can ignore the
constant term in $f$ and write (\ref{covmet}) as:
\begin{eqnarray}
\frac{1}{\alpha'}ds_{10}^2 & = &
        \left ( \frac{u^2}{\sqrt{Q}}\eta_{\mu\nu}dx^{\mu}dx^{\nu} +
        \frac{\sqrt{Q}}{u^2}du^2 \right ) +
        \sqrt{Q} \left ( d\gamma^2 + \cos^2\gamma d\phi^2 + \right.
\nonumber \\
 & &    \frac{1}{4}\sin^2\gamma d\theta^2 +
        \frac{1}{4}\sin^2\gamma \sin^2 \theta d\eta^2 +
\nonumber \\
 & & \left. \frac{1}{k^2} \sin^2\gamma
       \Big ( d\delta - \frac{k}{2}(\cos \theta - 1 )d\eta \Big )^2 \right ).
\label{IIBmetric}
\end{eqnarray}
where $Q$ is defined in terms of the IIB string coupling by:
\begin{equation}
Q = 4 \pi g_s k N.
\end{equation}
We will now show that the T-dual of our IIA metric is precisely this metric,
with $k$ identified with $n$, the number of NS5-branes. This
provides a non-trivial check of our solution and normalisations.

For the solution (\ref{metric}) the T-duality rules are fairly simple since
the metric is diagonal and only a few gauge fields are non-trivial.
 From \cite{Buscher1, Buscher2, Bergshoeff:1995as} we have the following
transformation rules for T-duality along the $\Phi$ direction:
\begin{eqnarray}
g_{\Phi \Phi} & = & g_{\Phi \Phi}^{-1} \\
g_{\Phi \sigma} & = & B_{\Phi \sigma} g_{\Phi \Phi}^{-1} \\
g_{\sigma \sigma} & = & g_{\sigma \sigma} + (B_{\Phi \sigma})^2 g_{\Phi \Phi}^{-1} \\
g_{MN} & = & g_{MN} \\
\exp(\phi) & = & \exp(\phi) g_{\Phi \Phi}^{-\frac{1}{2}} \label{T_dilaton} \\
F^{(5)}_{\Phi \eta \theta \omega \sigma} & = &
	 -\frac{3}{10}F_{\eta \theta \omega \sigma}
\end{eqnarray}
where all quantities on the left are the type IIB parameters, determined by
the type IIA parameters on the right. The indices $M, N$ are any spacetime
indices except $\Phi$ and $\sigma$. $F^{(5)}$ is the 5-form field strength
in type IIB and self-duality must be imposed. The transformations are written
for $\alpha ' = 1$.
It is now a simple matter to apply the T-duality rules and we find the type IIB
metric:
\begin{eqnarray}
\frac{1}{\alpha'}ds_{10}^2 & = &
        \left ( \frac{u^2}{\sqrt{Q}}\eta_{\mu\nu}dx^{\mu}dx^{\nu} +
        \frac{\sqrt{Q}}{u^2}du^2 \right ) +
        \left ( \sqrt{Q}d\eta^2 + \sqrt{Q} \sin^2\eta d\theta^2 + \right.
\nonumber \\
 & &	\frac{\sqrt{Q}}{4}\cos^2\eta d\omega^2 +
	\frac{\sqrt{Q}}{4}\cos^2\eta \sin^2 \omega d\sigma^2 +
\nonumber \\
 & & \left. \frac{\sqrt{Q}\cos^2\eta}{n^2}
       \Big ( d\Phi -c\frac{n}{2}(\cos \omega + c_0 )d\sigma \Big )^2 \right ).
\end{eqnarray}

The IIB string coupling is given in terms of the IIA string coupling by
equation~(\ref{T_dilaton}). Our IIA metric is only valid in the
near-horizon limit so we can't substitute $g_{\Phi \Phi}$ into
equation~(\ref{T_dilaton}) to relate the string couplings which are defined
asymptotically. However we have normalised our solution so that the metric
should be asymptotically flat. So asymptotically we must have
$g_{\Phi \Phi} = 1/c_1^2$ since the radius of the $\Phi$ direction is
$\sqrt{\alpha '}/c_1$.
So the string couplings are
related by:
\begin{equation}
g_s^{IIB} = c_1 g_s^{IIA}
\end{equation}
and so in terms of IIB quantities we have:
\begin{equation}
Q = 4 \pi g_s nN
\end{equation}

So we see that with $c_0 = -1$ we reproduce the metric for
$n$ D3-branes at a $\CC^2 / \ZZ_n$ orbifold fixed point.
The relation between the variables in equation~(\ref{IIBmetric}) and our
notation is easily seen to be: $\sin \eta = \cos \gamma$, $\theta = \phi$,
$\omega = \theta$, $c\sigma = \eta$ and $\Phi = \delta$.

Finally we give the non-trivial type IIB 5-form components:
\begin{equation}
(*F)_{\Phi \eta \theta \omega \sigma} = F_{\Phi \eta \theta \omega \sigma} =
   \frac{3 \pi c}{5 c_1} g_s N (\alpha ')^2 \sin \eta \cos^3 \eta \sin \omega
\end{equation}
        
\section{Conclusions}

We have shown the general form of the 11-dimensional supergravity solution
describing a 5-brane configuration with worldvolume $\RR^4 \times \Sigma$
with $\Sigma$ being a holomorphic 2-cycle in $\CC^2$.
This corresponds to a single 5-brane or multiple 5-branes depending on the
form of $\Sigma$, i.e. whether or not $\Sigma$ is simply connected. We have
also shown the reduction of this solution to 10 dimensions by
compactifying
one real coordinate of $\CC^2$ on a circle. This configuration will
generically correspond to a system of D4- and NS5-branes in type IIA string
theory.

We were unable to fully solve the source equations~(\ref{source}) but in the
special case we were interested in we could find a near-horizon solution for
N fully localised 5-branes intersecting n 5-branes which were smeared along the
localised 5-branes. Apart from the smearing, this reduces to a Hanany-Witten
configuration in type IIA describing the ${\cal N} = 2$ conformal field theory
with gauge group SU($N$)$^n$. Taking the near-horizon limit in
supergravity should give the Maldacena dual of the CFT and indeed we see that
this is plausible since we showed that the space factorises into the form
$AdS_5 \times M$ where $M$ is a compact manifold with the isometries
of the field theory R-symmetry group in addition to a U(1) due to
smearing.

There are a number of directions one can pursue.  One is to construct
the M-theory solution for non-conformal field theories in which the
corresponding type IIA picture has non-aligned D4-branes.  
Of interest also in the conformal theories considered here is to fully
localise the
NS5-branes. This would allow arbitrary gauge couplings for each SU($N$)
factor and would permit solutions with finite gauge couplings without
compactifying $x_6$. From the supergravity point of view it would also be
interesting to fully localise the M5-branes in 11 dimensions.
Another question \cite{danda2} which one can address is that of conformal field
theories with more exotic
matter content than fundamental and adjoint matter.  For example by
introducing a third NS5-brane and an orientifold 6-plane \cite{LL}
one can have symmetric or anti-symmetric second rank tensor
representations.

\vspace{0.5cm}
{\flushleft \bf \large Acknowledgements}
\vspace{0.2cm}

{\flushleft
We would like to thank
J. Maldacena, \"O. Sarioglu, D. Sorokin and C. Vafa for very helpful discussions.
AF is supported by NSF grant PHY-93-15811. DS is supported by the Deutsche
Forschungs Gemeinschaft.}

\vspace{0.5cm}
{\flushleft \bf \large Note added}
\vspace{0.2cm}

{\flushleft
A paper by D.~Youm \cite{Youm:1999zs} appeared in the archives on
the same day as this paper and another paper by A.~Loewy
\cite{Loewy:1999mn} 
appeared shortly after. Both these papers have some overlap with our
results in section~\ref{D4NS5solution}. 
}

\end{document}